\documentclass[conference]{IEEEtran}

\usepackage{cite}
\usepackage{float}
\usepackage{subcaption}
\usepackage{tabularx}
\usepackage{tcolorbox}
\usepackage{amsmath}
\usepackage{balance}
\usepackage{url}
\usepackage{caption}
\usepackage{hyperref}
\usepackage{colortbl}
\usepackage{xcolor}
\definecolor{lightgray}{gray}{0.9}

\usepackage{color}

\usepackage{soul}

\ifCLASSINFOpdf
  
\else
  
\fi

\hypersetup{pageanchor=true, hidelinks, linkcolor=black, urlcolor=blue, citecolor=black}

\begin{document}

\tcbset{
    myboxstyle/.style={
        colback=gray!10, %
        colframe=gray!70, %
        fonttitle=\bfseries, %
        coltitle=black, %
        boxrule=0.5mm, %
        arc=4mm, %
        left=0.2mm, %
        right=1mm, %
        top=0.5mm, %
        bottom=1mm, %
        width=8.4cm, %
        enlarge left by=0mm, %
        title={}, %
        before=\vspace{3mm}, %
        after=\vspace{3mm} %
    }
}

\title{Mining a Decade of Event Impacts on Contributor Dynamics in Ethereum: A Longitudinal Study}

\author{
\IEEEauthorblockN{M.~Vaccargiu\textsuperscript{1,5}, S.~Aufiero\textsuperscript{2}, C.~Ba\textsuperscript{3}, S.~Bartolucci\textsuperscript{2}, R.~Clegg\textsuperscript{3}, D.~Graziotin\textsuperscript{4}, R.~Neykova\textsuperscript{5}, R.~Tonelli\textsuperscript{1}, G.~Destefanis\textsuperscript{5}}
\IEEEauthorblockA{
\textsuperscript{1}University of Cagliari, Italy\\
\{matteo.vaccargiu,roberto.tonelli\}@unica.it\\
\textsuperscript{2}University College London, UK\\
\{sabrina.aufiero.22, silvia.bartolucci\}@ucl.ac.uk\\
\textsuperscript{3}Queen Mary University of London, UK\\
\{c.ba,r.clegg\}@qmul.ac.uk\\
\textsuperscript{4}University of Hohenheim, Germany\\
graziotin@uni-hohenheim.de\\
\textsuperscript{5}Brunel University of London, UK\\
\{rumyana.neykova,giuseppe.destefanis\}@brunel.ac.uk
}
}

\maketitle

\begin{abstract}
We analyze developer activity across 10 major Ethereum repositories (totaling 129884 commits, 40550 issues) spanning 10 years to examine how events such as technical upgrades, market events, and community decisions impact development. Through statistical, survival, and network analyses, we find that technical events prompt increased activity before the event, followed by reduced commit rates afterwards, whereas market events lead to more reactive development. Core infrastructure repositories like Go-Ethereum exhibit faster issue resolution compared to developer tools, and technical events enhance core team collaboration. 
Our findings show how different types of events shape development dynamics, offering insights for project managers and developers in maintaining development momentum through major transitions. This work contributes to understanding the resilience of development communities and their adaptation to ecosystem changes.
\end{abstract}

\IEEEpeerreviewmaketitle

\section{Introduction}
\label{intro}

Open-source software (OSS) projects play an important role in the development and maintenance of numerous technological infrastructures. Understanding the dynamics of these projects, particularly during significant events, can provide insights into their resilience, adaptability, and overall health \cite{malgonde2023resilience}. 
In recent years, blockchain technologies have emerged as a new frontier for OSS, presenting unique challenges and opportunities for collaborative development that extend and redefine traditional OSS paradigms \cite{Berdik2021}. 
Blockchain's convergence with OSS principles has created a distinct ecosystem with several notable characteristics. Unlike traditional OSS projects often led by a -- possibly distributed -- core team or a foundation\cite{Oreg2008Exploring}, blockchain projects adopt decentralized governance models\cite{Liu2021A}, aligning with their underlying philosophy but introducing new complexities in decision-making. Many projects incorporate native cryptocurrencies or tokens, introducing direct economic incentives for contributors and significantly impacting contribution patterns\cite{Petryk2023Impact}.Consensus-critical aspects of blockchain development are further impacted by complex technical challenges while managing community expectations, security concerns, and regulatory pressures  \cite{Chu2019The,TOUFAILY2021103444}. Major events -- such as security incidents, protocol changes, political attention, or even ordinary social media backlash are amplified by intense public scrutiny and can dramatically impact the project's trajectory\cite{Iqbal2020Asymmetric, ortu2022cryptocurrency}.
Ethereum\footnote{Ethereum is a decentralized, open-source blockchain platform that enables the creation of smart contracts, decentralized applications (dApps) \cite{Aufiero2024,ibba2024mindthedapp}, generate and exchange NFTs (non-fungible tokens), and many others. 
Ethereum's codebase is publicly available under open-source licenses, encouraging global collaboration and modification by a diverse community of developers.} serves as an ideal case study for this research due to its position as the leading smart contract platform, its significant market capitalization, and its history of major network upgrades and community-driven changes. The project's scale, complexity, and diverse ecosystem of developers and users provide a rich environment for examining OSS dynamics in the blockchain context.
Ethereum has faced numerous of the above-mentioned events throughout its history, which have tested the resilience, adaptability, and cohesion of its development community and have shaped its development path \cite{li2023development}. The dynamics and the effects of these events have, to the best of our knowledge, not been studied and understood.

In our study, we focus on analyzing the impact of major Ethereum events\footnote{A major Ethereum event represents a significant milestone or occurrence that fundamentally impacts matters belonging to at least two of the following categories: (1) ecosystem infrastructure (encompassing protocol-level changes and technical modifications); (2) market or community (including significant market reactions and community-driven changes); (3) development trajectory (covering coordinated development efforts and ecosystem-wide adaptations). Changes that affect only a single aspect of the ecosystem or require limited coordination across the network (e.g., minor bug fixes) are not classified as major. See Sec. \ref{meth} for more details.} on the Ethereum ecosystem by examining a set of ten key repositories. These repositories cover different aspects of Ethereum’s infrastructure, including core client implementations, smart contract languages, development tools, and libraries that enable interaction with the blockchain. Together, these repositories form the backbone of Ethereum's ecosystem, with interdependencies that reflect the interconnected nature of the platform. Studying them provides a view of how different components of the ecosystem respond to major events, offering insights into the overall adaptability and resilience of Ethereum's OSS community.

The nature resulting from the convergence of OSS and blockchain technologies drives us to investigate the broader impacts of ten of the major Ethereum events (described in greater detail in Sec. \ref{meth}).
By examining how these events influence team organization, developer response rates, commit activities\cite{Soto2017Analyzing}, and overall project health across key repositories, we aim to study how the Ethereum community adapts to these changes and provide insights into the ecosystem’s development dynamics\cite{Wang2019Unveiling}.
Among these metrics, commit activity serves as an important indicator, referring to the frequency of commits made by developers to a repository, capturing the level of ongoing development work and engagement within the project. 
Understanding commit activity helps identify periods of heightened or reduced activity, revealing how teams respond to major events, such as protocol upgrades or security incidents, and informing decisions about resource allocation and workload planning \cite{10.1145/3510003.3510205}.

In our study, we pose the following research questions:
\newline
\textbf{RQ 1: What is the time window during which the effects of major Ethereum events on developer collaboration in the Ethereum ecosystem are observable, before the network of collaboration returns to its normal state?}
Since our goal is to understand how specific major events have impacted the Ethereum developer collaboration ecosystem, it is essential to clearly define the \textit{before} and \textit{after} of each event, determining the time window beyond which the effects of the event are no longer observable. Through temporal analysis, we empirically established a 90-day window as the typical duration for event impacts to manifest and stabilize.
\newline
\textbf{RQ 2. How do major Ethereum events impact commit activity across key repositories in the Ethereum ecosystem?}
By analyzing fluctuations in commit activity around major events, we can identify when developers are most actively engaged or when contributions decrease. This understanding helps reveal how development efforts are coordinated, whether on a planned or voluntary basis, and provides insights into how developers respond to major changes in the ecosystem \cite{8115619}.
\newline
\textbf{RQ 3. How do major Ethereum events impact issue resolution time?}
Changes in issue resolution time indicate the efficiency and responsiveness of the development team during critical periods. This question helps assess the operational challenges faced by the team and whether developer actions are more structured (planned/allocated) or occur on a best-effort/volunteer basis \cite{1205177}.
\newline
\textbf{RQ 4. How do developer collaboration networks change during major Ethereum events?}
Examining shifts in collaboration networks reveals how team dynamics, roles, and communication patterns adapt during major events. This question is important because it helps us understand the underlying social structure of the developer community, which supports or constrains their collective response to changes in the ecosystem. We provide a complete replication package including all datasets, analysis scripts, results  at  \href{https://figshare.com/articles/journal_contribution/_b_Mining_a_Decade_of_Event_Impacts_on_Contributor_Dynamics_in_Ethereum_A_Longitudinal_Study_b_-_MSR_2025/27629559/1}{\textbf{this link}} to support reproducibility and verification \cite{Vaccargiu2025}.

\section{Related Work}
\label{rel_work}

Our study provides a blockchain-specific analysis of developers' activities and network structures. As shown in the following, previous studies have analyzed aspects of community interaction, event impact and network dynamics. Our work, on the other hand, focuses on a set of major events analysing how these positively influence the collaborative and technical aspects of Etheruem projects, a particular example of blockchain OSS. 
\paragraph{Studies on OSS Development}
The analysis of GitHub issues and comments in OSS projects has been explored by Mumtaz et al. \cite{Mumtaz202261}, who introduce features like "assign issues to issue commenters" and analyze social dynamics within software teams. Our study extends this approach by focusing on blockchain-specific OSS projects, examining not just social interactions but also the technical contributions through commits, thereby providing a more holistic view of developer behavior. Similarly, Jamieson et al. \cite{10548732} study the decentralized web communities by considering commits and the impact of project values on team dynamics. Our work diverges by employing network analysis techniques to scrutinize the structure of developer interactions over time, offering a novel perspective on how values and events influence OSS development processes.
Santos et al. \cite{Santos2023611} focus on facilitating task selection in OSS projects through textual data analysis. Their approach to understanding developer interactions and sentiments mirrors our methodology, which also seeks to unravel the complexities of developer communication in blockchain projects. However, our study extends beyond textual analysis to incorporate statistical and network analysis techniques, providing a more multidimensional view of developer behavior.
\paragraph{OSS Blockchain-Specific Studies}
Das et al. \cite{Das2022211} conduct an empirical analysis of interactions on commits within blockchain software repositories. Our methodology differs in that we not only analyze commits but also integrate issue resolution patterns and developer network dynamics to understand the broader implications of major blockchain events. Chakraborty et al. \cite{Chakraborty2018} provide an overview of software engineering practices in blockchain projects through a survey approach. In contrast, our study applies quantitative analysis of actual project data to uncover how events specifically impact these practices.
Ortu et al. \cite{Ortu201932} provide a comparative statistical characterization of traditional software systems versus blockchain-oriented software projects, exploring potential differences using a set of ten software metrics. Our research complements their findings by digging deeper into the blockchain-specific dynamics, especially how significant events influence these metrics within OSS development environments. 

\begin{table*}[h!]
\centering
\caption{Overview of the selected Ethereum repositories, sorted by total activity.}
\label{tab:repo_stats}
\resizebox{\textwidth}{!}{%
\begin{tabular}{l l r r r r }
\hline
\textbf{Repository} & \textbf{Description} & \textbf{Commits} & \textbf{Issues} & \textbf{Comments} \\ \hline
\textit{MetaMask} &  A widely used browser extension for managing Ethereum wallets and interacting with dApps  & 21121 & 11248 & 92676 \\ 
\textit{Solidity}  &    The main programming language for writing smart contracts on Ethereum  & 24617  & 5984  & 49744 \\ 
\textit{Go-ethereum} &  The primary client for running Ethereum nodes, crucial for the core operation of the blockchain  & 15373 & 8071  & 56109 \\ 
\textit{Chainlink} & A decentralized oracle network that allows smart contracts to connect with real-world data, essential for many DeFi applications  & 24060 & 430   & 26184 \\ 
\textit{Truffle}  & A development framework for compiling, deploying, and testing smart contracts  & 16047 & 2926  & 18546 \\ 
\textit{Web3-js}  & Another library for blockchain interaction, widely used in building decentralized applications   & 4005  & 3888  & 21189 \\
\textit{Hardhat} &  Contains the specifications for Ethereum 2, focusing on the shift to proof-of-stake and scalability  & 10650 & 2548  & 15715 \\ 
\textit{OpenZeppelin} & Provides a library of secure and reusable smart contracts, widely used in the ecosystem for security and development best practices   & 3648  & 1901  & 14359 \\ 
\textit{Consensus-Specs} & Contains the specifications for Ethereum 2, focusing on the shift to proof-of-stake and scalability   & 9685  & 920   & 8164  \\ 
\textit{Ethers-js} & A lightweight library for interacting with the Ethereum blockchain, popular among developers for its simplicity  & 678  & 2634  & 13961 \\ \hline
\textbf{Total} & -  & 129884  &  40550 & 316647 \\ \hline
\end{tabular}%
}
\end{table*}

We build on their foundational work by applying a similar analytical rigor to more granular, event-driven data, enhancing our understanding of how specific blockchain events impact software development practices.
\paragraph{Network Analysis in Software Engineering}
Network analysis has been employed to study community structures and interactions within software projects \cite{bartolucci2020butterfly, ferretti2020ethereum, kleinberg2000small, la2023game, louridas2008power, lucchini2020code, potanin2005scale, valverde2003hierarchical}.
Ao et al. \cite{Ao2021Temporal} and Said et al. \cite{Said2021Detailed} utilize network science to explore the community dynamics on the Ethereum blockchain. Our research builds upon these studies by specifically focusing on how major events such as market crashes or protocol upgrades affect the temporal and structural dynamics of developer networks, employing novel statistical techniques such as motif analysis and survival analysis.
\paragraph{Event Impact Analysis}
Major events and their impacts on software development processes are highlighted by Treude et al. \cite{Treude2017Unusual}, who examine unusual activities in GitHub projects. Our study expands on this by systematically defining event windows and employing rigorous statistical methods to assess the impact of these events on multiple facets of OSS development in blockchain projects. Additionally, studies by Pejić et al. \cite{Pejić2023Analyzing} and Kapengut et al. \cite{Kapengut2022An} focus on the impacts of COVID-19 and the transition to proof of stake on GitHub activities and network concentration, respectively.

\section{Dataset}
\label{dataset}

This study examines a set of 10 repositories from the Ethereum ecosystem. These repositories were selected because they represent various core components of the platform, including client implementations, smart contract programming, developer tools, and libraries for interacting with the blockchain. The goal is to understand how different parts of the ecosystem respond to major Ethereum events.
The Selected Repositories are Go-Ethereum (Geth), MetaMask, Ethers.js, Web3.js, Truffle, Solidity, Chainlink, Hardhat, Consensus-Specs and OpenZeppelin Contracts, which are described in detail below and in Table \ref{tab:repo_stats}.
The selected repositories represent different functional components within the Ethereum ecosystem, each serving distinct purposes while maintaining operational independence. While there are theoretical interdependencies in terms of functionality - for instance, developer tools like Truffle and Hardhat need to adapt to changes in Go-Ethereum specifications - the actual development processes and teams operate independently. Although Truffle has been officially sunset, as announced by ConsenSys, its historical significance and the transition towards Hardhat justify its inclusion in our analysis. The transition period from Truffle to Hardhat provides valuable insights into how the community adapts to evolving tools, demonstrating how independent development teams respond to ecosystem-wide changes.

Updates to the Solidity programming language can impact tools like Ethers.js, Web3.js, and OpenZeppelin Contracts, which must stay compatible. Such interconnections mean changes in one area, like a protocol update, quickly ripple through and require adjustments across tools. 
MetaMask, a bridge between users and decentralized applications, relies on Web3.js and Ethers.js for Ethereum interactions. Changes in MetaMask or these libraries significantly impact user experience and development practices. Chainlink, providing external data to smart contracts, is also vital for projects requiring secure data feeds.
These interconnections mean that changes from major Ethereum events propagate through the ecosystem, impacting various repositories and the developer community. Examining these linked repositories provides insight into how significant changes influence Ethereum’s development and operation. This interconnected dataset enables a thorough analysis of Ethereum’s systemic responses, including transitional phases like Truffle to Hardhat.
Including repositories of varying sizes and activity levels captures a broad spectrum of responses within the Ethereum community. Larger repositories like MetaMask and Solidity represent core infrastructure, while smaller or transitioning ones, like Truffle’s shift to Hardhat, demonstrate adaptability. Analyzing both stable and evolving repositories reveals how different parts of the ecosystem respond, whether by maintaining stability or adapting to new technologies.

\section{Methodology}
\label{meth}

\begin{figure*}[ht]
  \centering
  \includegraphics[width=\textwidth]{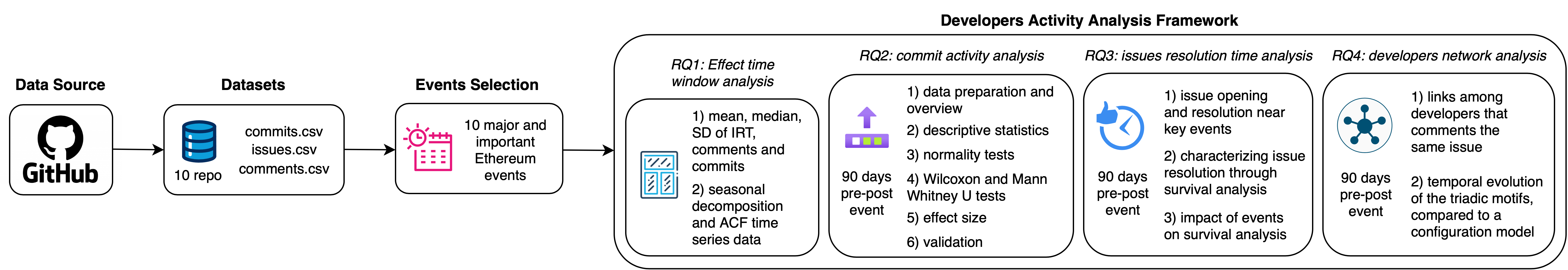}
  \caption{Description of the methodology}
  \label{fig:methodology}
    
\end{figure*}

\begin{table*}[h!]
\caption{Major Ethereum Events and Their Impact on Development Activity}
\centering
\resizebox{\textwidth}{!}{
\begin{tabular}{l l l l }
\hline
\textbf{Event} & \textbf{Date} & \textbf{Significance} & \textbf{Impact} \\ \hline
\textit{Frontier Release} & July 30, 2015 & First official Ethereum release, allowing mining and dApp development. & Increased development activity and community engagement. \\ 
\textit{The DAO Creation and Hack} & Jun 17, 2016 & Major project built on Ethereum, hacked leading to a hard fork. & Led to Ethereum (ETH) and Ethereum Classic (ETC) split, impacting governance and developer focus. \\ 
\textit{Cryptocurrency Boom} & Dec 17, 2017 & Ethereum's price peak during the broader cryptocurrency surge. & Raised mainstream interest, shifting developer priorities. \\ 
\textit{Market Crash (COVID-19)} & Mar 13, 2020 & Ethereum's market value dropped with the global downturn. & Sparked growth in DeFi, shifting focus to decentralized finance projects. \\ 
\textit{Beacon Chain Launch} & Dec 1, 2020 & Introduction of the proof-of-stake consensus mechanism. & Changed development approaches and resource allocation. \\ 
\textit{London Hard Fork} & Aug 5, 2021 & Implemented EIP-1559, restructuring Ethereum's fee system. & Required developers to adapt to new fee structures. \\ 
\textit{Arrow Glacier Update} & Dec 9, 2021 & Delayed the difficulty bomb, encouraging the proof-of-stake transition. & Maintained network functionality, impacting development timelines. \\ 
\textit{Ropsten Testnet Merge} & Jun 8, 2022 & Key test for the mainnet transition to proof-of-stake. & Provided insights and preparation for the mainnet merge. \\ 
\textit{The Merge} & Sep 15, 2022 & Transition of Ethereum mainnet to proof-of-stake. & Triggered significant updates and adaptations for developers. \\ 
\textit{Shanghai Upgrade} & Apr 12, 2023 & Enabled staked ETH withdrawals for Ethereum 2.0. & Improved key network functionality, influencing developer activity. \\ \hline
\end{tabular}
}
\label{tab:events}
\end{table*}

Our systematic investigation of developer activity in the Ethereum ecosystem follows a structured approach, from detailed data collection through multiple layers of analysis, as illustrated in Fig. \ref{fig:methodology}. This methodology enables a thorough examination of how major events influence development patterns across key repositories.

Starting with data collection from GitHub, shown in the left section of Fig. \ref{fig:methodology},  we identified and extracted data from 10 central Ethereum repositories, focusing on core infrastructure and widely-used development tools. For each repository, we systematically collected three primary datasets: commit histories with metadata, issue tracking information (including creation and resolution timestamps), and developer comments.

The second phase is event selection. We analyzed Ethereum’s history to identify 10 major events that have shaped the ecosystem, listed in Table \ref{tab:events}. These events, spanning from Ethereum’s launch to recent protocol updates, offer varied contexts for examining developer responses to distinct challenges and changes.

The identification and classification of major events followed a three-phase protocol. First, we constructed an event pool by aggregating data from authoritative sources: Ethereum Foundation communications, network upgrade histories, market data, and security incident reports during 2014-2024. Two authors then independently evaluated each event using three classification criteria. \textbf{Infrastructure Impact} related to network-wide protocol changes and technical modifications (e.g., \textit{The Merge}, \textit{London Hard Fork}). \textbf{Market/Community Impact} includes significant market reactions and community-driven changes (e.g., \textit{COVID-19 crash}, \textit{The DAO hack}). \textbf{Development Trajectory Impact} covers events requiring coordinated development efforts or establishing new patterns (e.g., \textit{Arrow Glacier Update}, \textit{Frontier Release}). Events meeting criteria from at least two categories were classified as major, ensuring broad ecosystem impact rather than isolated effects. For diverging classifications, a third author conducted an independent review. Event validation involved examining Ethereum development community archives (GitHub discussions, Improvement Proposals, official announcements), yielding our final set of 10 major events.

The events span from the Frontier Release (July 30, 2015) to the Shanghai Upgrade (April 12, 2023), covering technical milestones like The Merge, market events such as the COVID-19 crash, and community-driven incidents like The DAO hack. The DAO Creation which took place on April 12, 2016 and the DAO Hack which took place on June 17, 2016 are two separate but very close events (less than 90 days apart). As explicated below, we avoided overlapping effects for streamlining our analysis; therefore, we took the date of the hack as the reference date. Table \ref{tab:events} provides details on each event’s significance and its impact on development activity.

The core of our analysis examines four distinct aspects of developer activity, each aligned with a specific research question.

To answer RQ1, we analyzed three key metrics across all repositories. The first metric, \textbf{Issue Resolution Time}, measures the time between issue creation and closure. The second metric, \textbf{Monthly Comment Count}, reflects the intensity of collaboration. Lastly, \textbf{Monthly Commit Count} serves as an indicator of development activity.

Based on initial observations and statistical analysis, we establish a \textbf{Pre-Event Window} covering the 90 days before each event to set baseline activity levels, and a \textbf{Post-Event Window} covering the 90 days following the event to track immediate and residual effects. For each metric, we conduct seasonal decomposition, partitioning the data into three components: \textbf{Trend}, which captures long-term movement patterns; \textbf{Seasonality}, which reflects regular, recurring fluctuations; and \textbf{Residuals}, representing remaining variation after accounting for trend and seasonality.

To validate these components, we use autocorrelation function (ACF) analysis, assessing correlations between time series values at different lags to ensure robustness in our decomposition and interpretation.

To answer RQ2, we conducted a detailed analysis of commit patterns. Our first step focused on data preparation and overview, where we normalized commit counts to enable fair comparisons across repositories with varying activity levels.
Next, we performed descriptive statistical analysis, calculating the means, medians, and standard deviations of commit activity for both pre-event and post-event periods. To understand the distribution characteristics of our data, we examined histograms, Q-Q plots, and conducted Shapiro-Wilk tests\cite{shapiro1965analysis}.

For our inferential statistical analysis, we examined each repository-event pair by testing the following hypothesis: \textbf{$H_0$:} \textit{The distribution of commit counts in the 90 days before the event is the same as the distribution in the 90 days after the event} (with \textbf{$H_A$} stating that the distribution differs). Given that each repository serves different purposes in the Ethereum ecosystem and operates independently, we treated them as separate experiment groups. For each repository, we conducted 10 tests (one per event). Since repositories are independent, for each repository we applied the Benjamini-Hochberg procedure \cite{Thissen2002Quick} to control the False Discovery Rate (FDR) at $\alpha = .05$ across its 10 events, which is suitable for balancing Type I and Type II errors \cite{Keselman2002Controlling} in studies akin to ours.

To quantify the magnitude of changes, we calculated effect sizes using the Z-score through the formula $r = Z / \sqrt{N}$, where $Z$ represents the standardized test statistic and $N$ is the number of observations. We interpreted these values using established thresholds: $>.10$ for small effects, $>.30$ for moderate effects, and $>.50$ for large effects.

Finally, to validate that our results were not due to random fluctuations, we performed a control analysis by testing against 100 randomly selected events and reshuffling time series.

For RQ3, We employed survival analysis techniques to examine issue management patterns. First, issue lifecycles are tracked by visualizing resolved issues at their closure points, while unresolved issues are observed until the study’s end. To account for both resolved and unresolved (censored) issues, Kaplan-Meier estimation is used, producing survival curves that represent resolution probabilities over time. Event impacts are then assessed through log-rank tests, comparing survival distributions before and after events (against the null hypothesis: \textbf{$H_0$:} \textit{H0: Events do not change how quickly issues get resolved}), with a significance level of $\alpha = .05$, and applying the Benjamini-Hochberg procedure to control the FDR for consistency with RQ2.

The final component of our methodology, for answering RQ4, examines collaboration networks. Developer interactions are represented by linking individuals who comment on the same issues, with the strength of each connection based on the number of shared issues. To observe network evolution, temporal analysis is performed over twelve-month periods segmented into three phases: a baseline phase from six to three months pre-event, an impact period spanning three months before and after the event, and a stabilization phase covering three to six months post-event.

Motif analysis identifies recurring interaction patterns, using the Configuration Model as a null model and Z-scores to quantify the significance of these patterns. Consistent 90-day windows established in RQ1 are applied throughout all analyses, ensuring comparable metrics across repositories. This methodology provides a multi-dimensional view of event impacts, examining individual contributions, issue resolution efficiency, and collaborative dynamics within the ecosystem.

Two events in our dataset—London Hard Fork/Arrow Glacier Update and Ropsten Testnet Merge/The Merge—have overlapping post-event and pre-event periods, where the three-month window of one event extends into the three-month period of another. We maintained a consistent 90-day analysis period for all events, based on the empirical findings from RQ1. Our analysis examines the relative changes in developer activity surrounding each event, with the validation against random events in RQ2 confirming the observed patterns.

\begin{table}[h!]
\caption{Metrics across repositories}
\resizebox{\columnwidth}{!}{%
\begin{tabular}{c cc cc cc }
\hline
\multicolumn{1}{l}{
\textbf{Repository}}
  &
  \multicolumn{2}{c}{\textbf{Issue resol. time}} &
  \multicolumn{2}{c}{\textbf{N° of Comments}} &
  \multicolumn{2}{c}{\textbf{N° of Commits}} \\ 
 &
  \multicolumn{1}{r}{Mean} &
  \multicolumn{1}{r}{Median} &
  \multicolumn{1}{r}{Mean} &
  \multicolumn{1}{r}{Median} &
  \multicolumn{1}{r}{Mean} &
  \multicolumn{1}{r}{Median} \\ \hline
\multicolumn{1}{l}{\textbf{Overall}} & \multicolumn{1}{r}{143.8} & \multicolumn{1}{r}{16.3} & \multicolumn{1}{r}{2473.8} & \multicolumn{1}{r}{2854.0} & \multicolumn{1}{r}{1006.8} & \multicolumn{1}{r}{1078.0} \\
\multicolumn{1}{l}{\textit{MetaMask}} &
  \multicolumn{1}{r}{146.9} &
  \multicolumn{1}{r}{23.9} &
  \multicolumn{1}{r}{891.1} &
  \multicolumn{1}{r}{733.0} &
  \multicolumn{1}{r}{193.8} &
  \multicolumn{1}{r}{171.0} \\
\multicolumn{1}{l}{\textit{Solidity}} &
  \multicolumn{1}{r}{254.6} &
  \multicolumn{1}{r}{33.2} &
  \multicolumn{1}{r}{456.4} &
  \multicolumn{1}{r}{453.0} &
  \multicolumn{1}{r}{192.3} &
  \multicolumn{1}{r}{176.5} \\
\multicolumn{1}{l}
{\textit{Go-ethereum}} &
  \multicolumn{1}{r}{122.5} &
  \multicolumn{1}{r}{5.4} &
  \multicolumn{1}{r}{438.3} &
  \multicolumn{1}{r}{441.0} &
  \multicolumn{1}{r}{119.2} &
  \multicolumn{1}{r}{82.0} \\
\multicolumn{1}{l} {\textit{Chainlink}} &
  \multicolumn{1}{r}{85.8} &
  \multicolumn{1}{r}{12.7} &
  \multicolumn{1}{r}{323.3} &
  \multicolumn{1}{r}{220.0} &
  \multicolumn{1}{r}{293.4} &
  \multicolumn{1}{r}{278.0} \\
\multicolumn{1}{l}{\textit{Truffle}} &
  \multicolumn{1}{r}{116.9} &
  \multicolumn{1}{r}{24.2} &
  \multicolumn{1}{r}{178.3} &
  \multicolumn{1}{r}{175.5} &
  \multicolumn{1}{r}{160.5} &
  \multicolumn{1}{r}{141.0} \\
  \multicolumn{1}{l} {\textit{Web3-js}} &
  \multicolumn{1}{r}{127.8} &
  \multicolumn{1}{r}{29.8} &
  \multicolumn{1}{r}{178.1} &
  \multicolumn{1}{r}{165.0} &
  \multicolumn{1}{r}{33.4} &
  \multicolumn{1}{r}{20.0} \\
\multicolumn{1}{l}{\textit{Hardhat}} &
  \multicolumn{1}{r}{103.9} &
  \multicolumn{1}{r}{17.0} &
  \multicolumn{1}{r}{206.8} &
  \multicolumn{1}{r}{190.0} &
  \multicolumn{1}{r}{138.3} &
  \multicolumn{1}{r}{123.0} \\ 
\multicolumn{1}{l}{\textit{OpenZeppelin}} &
  \multicolumn{1}{r}{109.8} &
  \multicolumn{1}{r}{7.3} &
  \multicolumn{1}{r}{149.6} &
  \multicolumn{1}{r}{139.0} &
  \multicolumn{1}{r}{37.6} &
  \multicolumn{1}{r}{31.0} \\ 
\multicolumn{1}{l}{\textit{Consensus-Specs}}  & \multicolumn{1}{r}{105.3} & \multicolumn{1}{r}{14.0} & \multicolumn{1}{r}{113.4}  & \multicolumn{1}{r}{70.5}   & \multicolumn{1}{r}{134.5}  & \multicolumn{1}{r}{98.5}   \\ 
\multicolumn{1}{l}{\textit{Ethers-js}} &
  \multicolumn{1}{r}{69.1} &
  \multicolumn{1}{r}{6.0} &
  \multicolumn{1}{r}{143.9} &
  \multicolumn{1}{r}{145.0} &
  \multicolumn{1}{r}{28.2} &
  \multicolumn{1}{r}{26.0} \\ 
 \hline
\end{tabular}%
}
\label{tab:metriche}
\end{table}

\begin{figure}[t!]   
    \centering \includegraphics[scale=0.3]{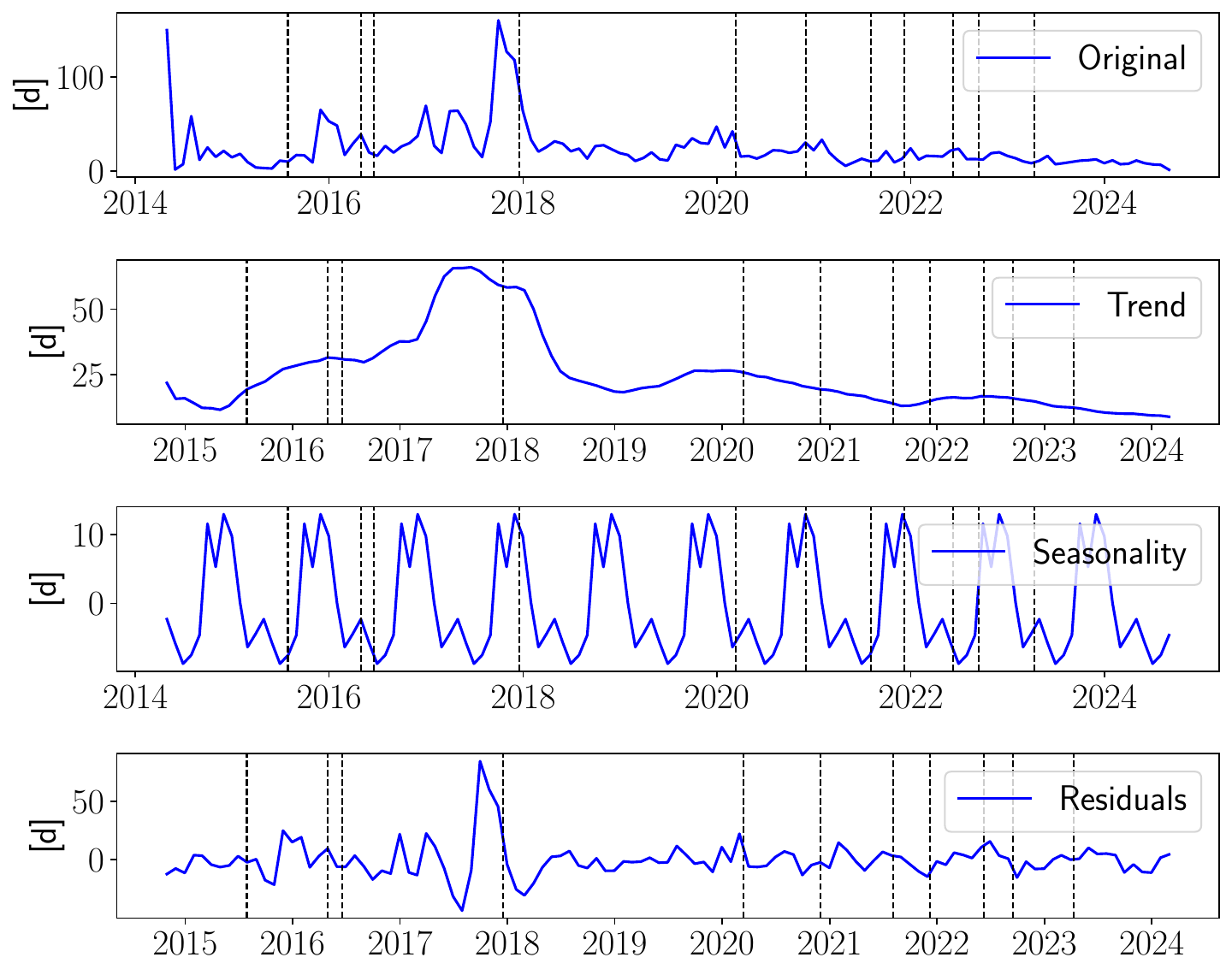}
    \caption{Analysis for the resolution times of issues}
    \label{fig: season resol time}
\end{figure}

\begin{figure}[t!]
\centering
    \centering \includegraphics[scale=0.3]{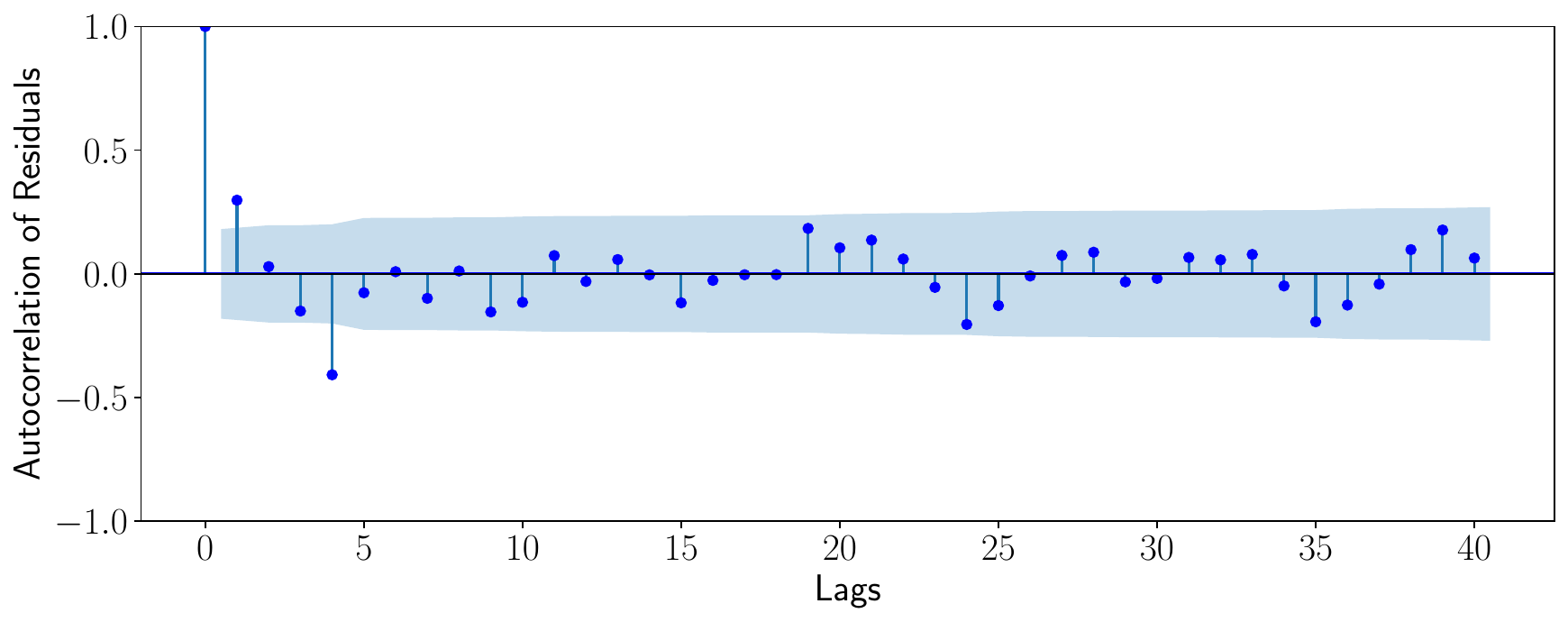} 
    \caption{ACF function for commits} \label{fig: ACF commits}
\end{figure}

\section{Analyzing the Temporal Impact of Significant Events on Developer Collaboration}\label{RQ1}

To understand the impact of significant events on developer collaboration, we define clear \textit{before} and \textit{after} periods for each event. This allows us to identify the time window beyond which the effects of the event are no longer noticeable.

\textbf{Issue resolution time}, defined as the time between an issue's creation and closure \cite{ortu2015bullies} ($M = 143$, $SD = 288$, $median = 16$ days). This distribution is skewed, with most issues resolved quickly and a smaller subset remaining unresolved for extended periods.\\
\textbf{Number of comments created monthly} ($M = 2473$, $SD = 1488$, $median = 2854$ comments).\\
\textbf{Number of commits created monthly} ($M = 1006$, $SD = 463$, $median = 1078$ commits)

For each of these metrics, we compute the overall mean and median values, as well as for each individual repository, as summarized in Table \ref{tab:metriche}. The distribution of issue resolution times is consistently skewed, with the median significantly lower than the mean across all repositories. In contrast, the number of comments and commits show closer alignment between the mean and median, indicating more balanced distributions for these metrics.
To analyze the temporal patterns in these metrics, we use the median values of resolution times, comments created, and commits on a monthly basis. This approach helps account for any potential seasonality in the data. We selected the median over the mean due to the skewed distribution of resolution times, as the median provides a more robust measure of central tendency and is less affected by outliers or extreme values. 

Fig. \ref{fig: season resol time} presents the seasonal decomposition of issue resolution times into trend, seasonality, and residuals. The original time series (top panel) shows raw data from 2015-2024, with notable variations in median resolution times. The trend component (second panel) reveals long-term patterns, while the seasonal component (third panel) captures recurring cycles in resolution times throughout each year. The residuals (bottom panel) show the remaining variance after accounting for both trend and seasonal effects, with their random distribution suggesting the model has effectively captured the main temporal patterns.
The bottom section displays the residuals, which represent the remaining variance after accounting for both trend and seasonality. The residuals reflect random fluctuations in the data that are not explained by the long-term trend or seasonal patterns. The random distribution of residuals suggests that the model has captured most of the meaningful variations.
To confirm the randomness of the residuals, we compute the autocorrelation function (ACF). The ACF measures how a time series is correlated with its past values at different time lags, helping us understand the relationship between an observation at time $t$ and earlier observations at times $t - 1$, $t - 2$, and so on. Ideally, the residuals should resemble white noise — random, without discernible patterns — indicating that the model (trend and seasonality) has adequately captured the time series dynamics. If the residuals exhibit significant autocorrelation, the model has not fully captured the underlying patterns, and some structure remains.

We compute the autocorrelation function (ACF) analysis for each metric - resolution times, comments, and commits. For resolution times, we observe an inverse relationship at lag 3, while comment patterns show no statistically significant autocorrelation. 
Fig. \ref{fig: ACF commits} shows the ACF analysis for commits, where bars outside the blue confidence interval indicate statistically significant temporal correlations. The analysis reveals significant lags at 1 (positive) and 4 (negative), indicating that high commit activity in one month predicts similar activity the next month, but tends to decrease after four months. The positive autocorrelation at lag 1 suggests persistent complexity, where certain issues span multiple days or indicate development process bottlenecks. In contrast, the negative autocorrelation at lag 4 suggests that teams typically resolve longer issues or adapt processes within this timeframe, leading to decreased commit activity.
Given our dataset of 129884 commits and 40550 issues, commits provide the most detailed measure of developer activity and are our primary focus. Through this ACF analysis, we define the \textit{post-event window} as the third month after an event, when the final effects on commits are observed, with normal activity resuming by the fourth month. For consistency, we consider the 90 days before an event the \textit{pre-event} and the 90 days immediately following as the \textit{post-event}.

\begin{tcolorbox}[right=0.1cm,left=0.1cm,top=0.1cm,bottom=0.1cm]
\textbf{Answer to RQ1}: The effects of major Ethereum events on developer collaboration are observable for 90 days (approximately three months) after the event. By the fourth month, activity levels typically stabilize. 
\end{tcolorbox}

\section{Impact Analysis of Major Ethereum Events on Developer Commit Activity}\label{RQ2}

Understanding how major Ethereum events impact development activity is important for assessing how developer teams respond to network changes. By tracking commit activity before and after these events, we can determine whether development efforts are planned or reactive. This analysis highlights periods of increased or decreased engagement, offering a clearer view of how updates or disruptions affect the ecosystem’s overall activity.

We observed significant variations in commit activity across repositories in response to major Ethereum events. By comparing the number of commits before and after each event, we can identify whether developers tend to be more active before or after these occurrences.

For the Frontier Release, we observed that the number of commits generally decreases post-event in Go-ethereum, Solidity, and Web3.js, while MetaMask Extension remains stable. The DAO Creation and Hack led to an increase in commits for Solidity and MetaMask Extension, indicating post-event code updates. The COVID-19 Crash resulted in more commits for Go-ethereum, Hardhat, and MetaMask Extension. Notably, Hardhat showed significant increases in commit activity after the Beacon Chain Launch, The Merge, and Shanghai Upgrade.

\begin{figure}
    \centering
    \includegraphics[width=0.9\linewidth]{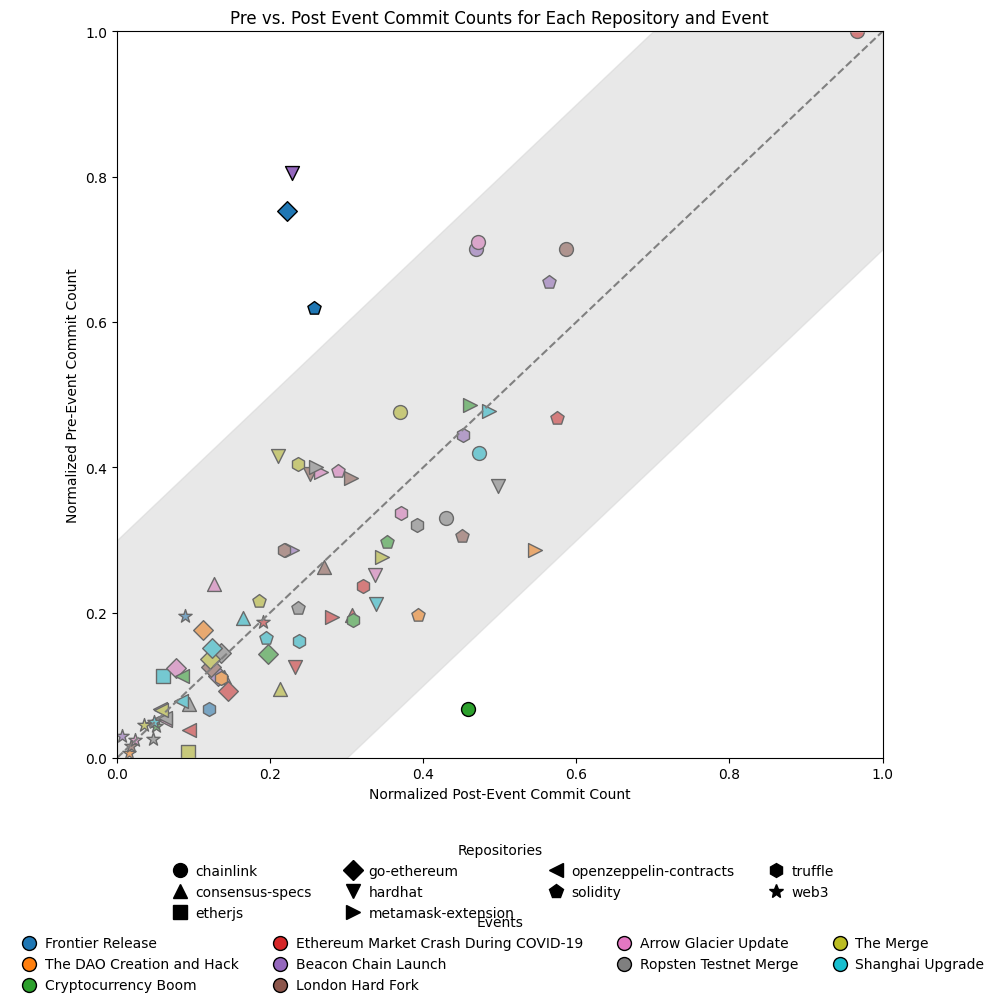}
    \caption{Pre vs. Post event commit counts for each repository and event}
    \label{fig:PrePostNormCommitCount}
\end{figure}

Fig. \ref{fig:PrePostNormCommitCount} highlights the changes in pre- and post-event activity that fall outside the gray confidence area of $r-value = .30$. The only event with a significant increase in commits is the Cryptocurrency Boom for Chainlink. Conversely, there is a notable decrease in commits for the Frontier Release in Solidity and Go-ethereum, and for the Beacon Chain Launch in Hardhat.

\begin{figure}
    \centering
    \includegraphics[width=0.9\linewidth]{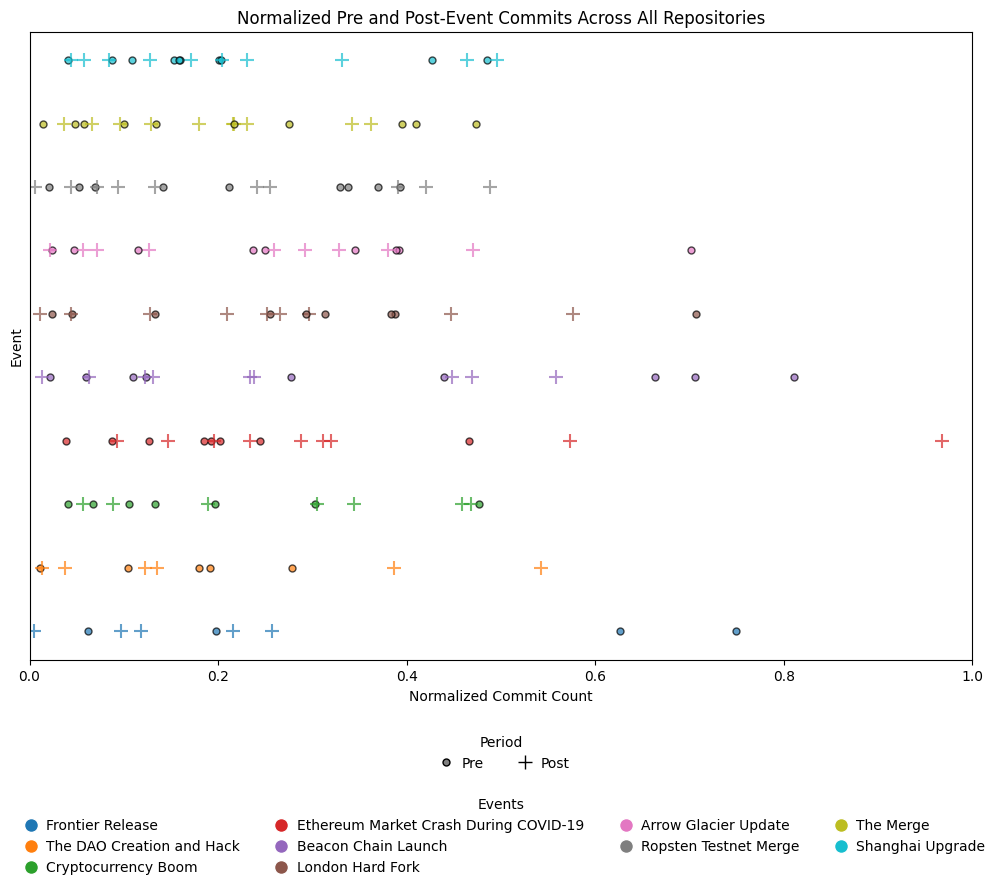}
    \caption{Normalized count pre (circle) and post (cross) event commits across all repositories}
    \label{fig:EventPrePostCount}
\end{figure}

Fig.~\ref{fig:EventPrePostCount} reveals varying impacts of events on commit activity. Market-driven events (COVID-19 Crash, DAO Creation) trigger post-event commit peaks, while planned technical changes (Beacon Chain Launch, The Merge, Frontier Release) show pre-event activity concentration, suggesting proactive development. Technical milestones tend to slow subsequent commit activity during protocol adaptation, whereas market events stimulate reactive development responding to economic shifts.
Repository responses varied by type: infrastructure repositories (Go-ethereum, Solidity) concentrate activity around core updates; development tools (Hardhat, Truffle) show post-upgrade compatibility spikes; market-sensitive repositories (Chainlink, MetaMask) demonstrate rapid responses to external pressures; and libraries (Web3.js, Ethers.js) maintain consistent activity with targeted increases during technical transitions.
Shapiro-Wilk tests ($p < .05$) showed non-normal distributions for all repository-event pairs, necessitating the use of non-parametric methods.

We, hence, employed the Wilcoxon signed-rank test for our Hypothesis on comparing pre- and post-event commit activity within the same repository, as samples are paired.
\begin{table*}[t]
\rowcolors{2}{gray!15}{white}
\footnotesize
\caption{Significant Results for Wilcoxon Tests after Benjamini-Hochberg Correction ($p < .05$)}
\resizebox{\textwidth}{!}{
\begin{tabular}{lcccccccccccc}
\hline
\textit{\textbf{Repository} } & \textbf{1 - Frontier} & \textbf{2 - DAO Hack} &  \textbf{3 - Crypto Boom}  & \textbf{4 - COVID Crash}& \textbf{5 - Beacon} & \textbf{6 - London} & \textbf{7 - Arrow Gl.} & \textbf{8 - Ropsten} & \textbf{9 - Merge} & \textbf{10 - Shanghai}\\
\hline
\textit{MetaMask} &  & x &  & x &  &  & x &  &  &  \\
\textit{Solidity} & x & x &  &  &  & x &  &  &  &  \\
\textit{Go-ethereum} & x &  &  & x &  &  & x &  &  &  \\
\textit{Chainlink} &  &  &  &  & x &  & x &  &  &  \\
\textit{Truffle} &  &  & x &  &  & x &  &  & x &  \\
\textit{Web3.js} & x &  &  &  & x &  &  &  &  &  \\
\textit{Hardhat} &  &  &  & x & x & x &  &  & x & x \\
\textit{OpenZeppelin} &  &  &  &  &  &  &  &  &  &  \\
\textit{Consensus-Specs} &  &  &  &  &  &  & x &  & x &  \\
\textit{Ethers.js} &  &  &  &  &  &  &  &  &  &  \\
\hline
\end{tabular}
}
\label{tab:wilcoxon-results}
\end{table*}

Table~\ref{tab:wilcoxon-results} shows significant results ($p < .05$) from the Wilcoxon tests after applying the Benjamini-Hochberg correction. Core repositories (Go-ethereum, Solidity) show significant changes during foundational events like the Frontier Release, while development tools (Hardhat, Truffle) display significant activity around major upgrades, indicating efforts to ensure compatibility. MetaMask exhibits heightened activity around security incidents and market crashes.
We calculated effect sizes (r-values) for the Wilcoxon signed-rank test (the complete effect size values list is available in our replication package).
Negative r-values were common, indicating a general decrease in commit activity after major events, suggesting a period of adjustment. The main decreases with ($r = -.86$) are for in the Web3-js repository for The DAO Creation and Hack, Beacon Chain Launch and London Hard Fork events. Further major changes are for The Merge ($r = -.83$) and Shanghai Upgrade ($r = -.82$) in Ethers-js and Frontier Release ($r = -.83$) for Go-ethereum and for solidity. 
Finally, market-related events such as Ethereum Market Crash During COVID-19 ($r = -.82$) and Beacon Chain Launch ($r = -.83$) have an interesting impact for OpenZeppelin. 

To validate our findings, we repeated the same analysis using 100 random events and reshuffled time series for the repositories. The analysis yielded 0 significant results after correction, compared to the observed significant events in Table~\ref{tab:wilcoxon-results}. This confirms that the observed effects are due to the selected major Ethereum events and not random chance.

\begin{tcolorbox}[right=0.1cm,left=0.1cm,top=0.1cm,bottom=0.1cm]
\textbf{Answer to RQ2:} 
Major Ethereum events significantly impact commit activity across key repositories, with the most significant events being the Frontier Release, The DAO Creation and Hack, and the Beacon Chain Launch. Core repositories show increased activity around foundational and security events, while development tools respond to major upgrades. 
\end{tcolorbox}

\section{Impact of Major Ethereum Events on Issue Resolution Time}\label{RQ3}

To evaluate whether development efforts are planned or reactive to major events, we analyzed issue activity before and after each event. Daily issue data was collected from each repository to observe trends in issue openings and closures, focusing on shifts in resolution patterns. Kaplan-Meier estimation was used to model resolution time distributions, with event dates serving as covariates to compare resolution behavior before and after each event. Using the list of the major events from RQ2, we visualized changes in issue activity within the 1-month and 3-month timeframes identified in previous sections. Notable time-based fluctuations in issue activity across repositories highlight the importance of dynamically analyzing these patterns to capture the temporal impact of major events on resolution times.

\begin{figure}
    \centering
    \includegraphics[width=\linewidth]{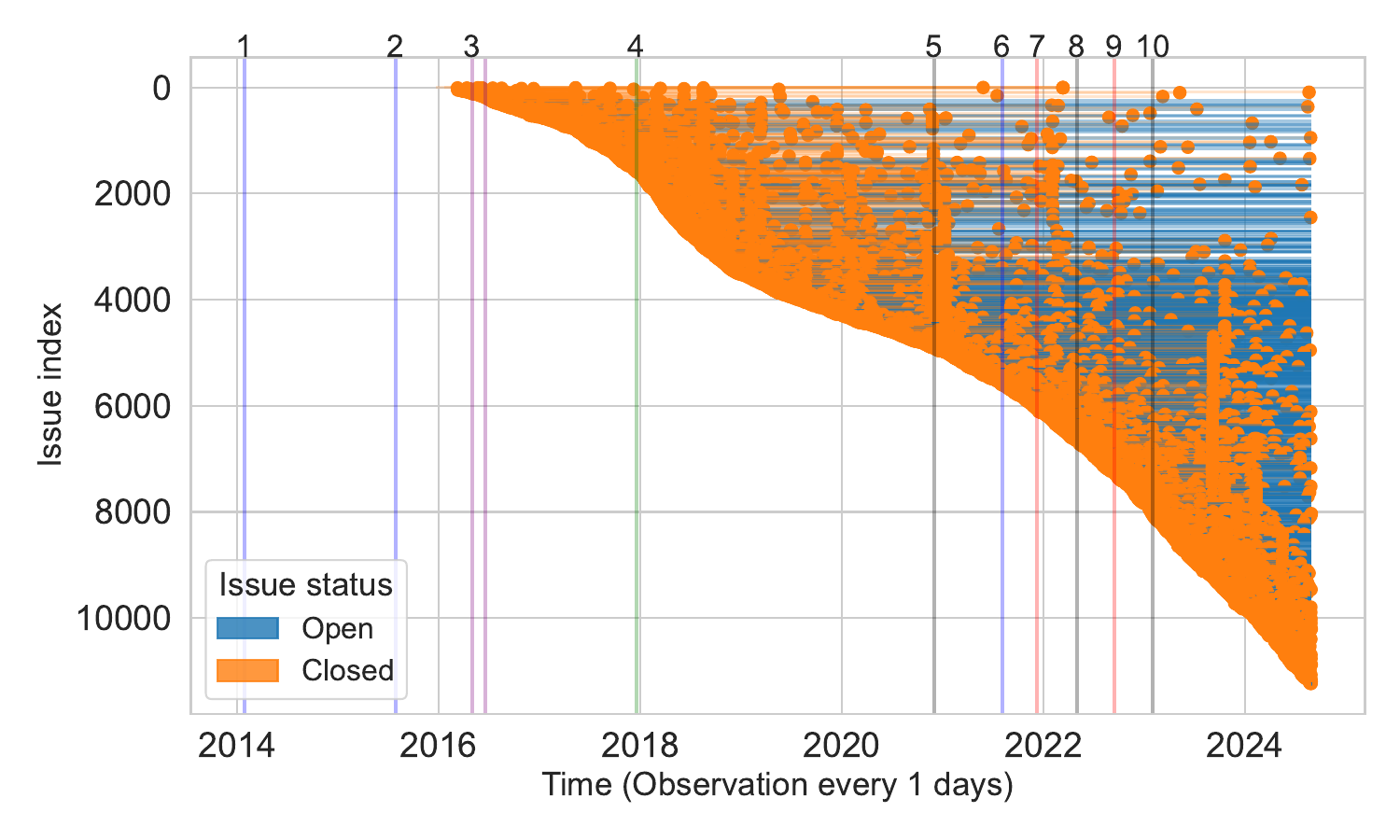}
    \caption{MetaMask's issue resolution and events}
    \label{fig:lifelines-metamask}
\end{figure}

\paragraph{Issue Opening and Resolution Near Key Events}
To examine how issue activity aligns with major Ethereum events, we plotted issue resolution times across all repositories, with each line representing an issue. Resolved issues are marked with an orange dot at the end, while unresolved issues extend in blue to the end of the observation period. Sorting issues by creation date allows us to track resolution patterns before and after specific events. Here, only the plot for MetaMask is shown (Fig.~\ref{fig:lifelines-metamask}).
Across repositories, a common trend is the spike in issue resolutions around major events. In MetaMask, we observe notable resolution activity around the Beacon Chain Launch (event 5) and the Arrow Glacier Update (event 7). Similar patterns are seen in other repositories. For instance, Go-Ethereum shows increased resolution activity near key events and an additional spike in older issue resolutions before 2019. Solidity also displays surges around the Beacon Chain Launch (event 5), The Merge (event 9), and the Shanghai Upgrade (event 10). 

For the other repositories, similar resolution spikes are evident. For example, Web3.js, Hardhat, and Ether.js show increased resolution activity around the Shanghai Upgrade (event 10), while the Arrow Glacier Update (event 7) coincides with significant resolution efforts in Ether.js, OpenZeppelin, and Chainlink. The Merge (event 9) also influences activity in Hardhat and Chainlink.
While not all resolution spikes directly correspond to major events—some relate to each repository’s unique milestones—these findings illustrate a general pattern of heightened issue resolution, often in anticipation of or response to major Ethereum events. 

\begin{figure}
    \centering
    \includegraphics[width=\linewidth]{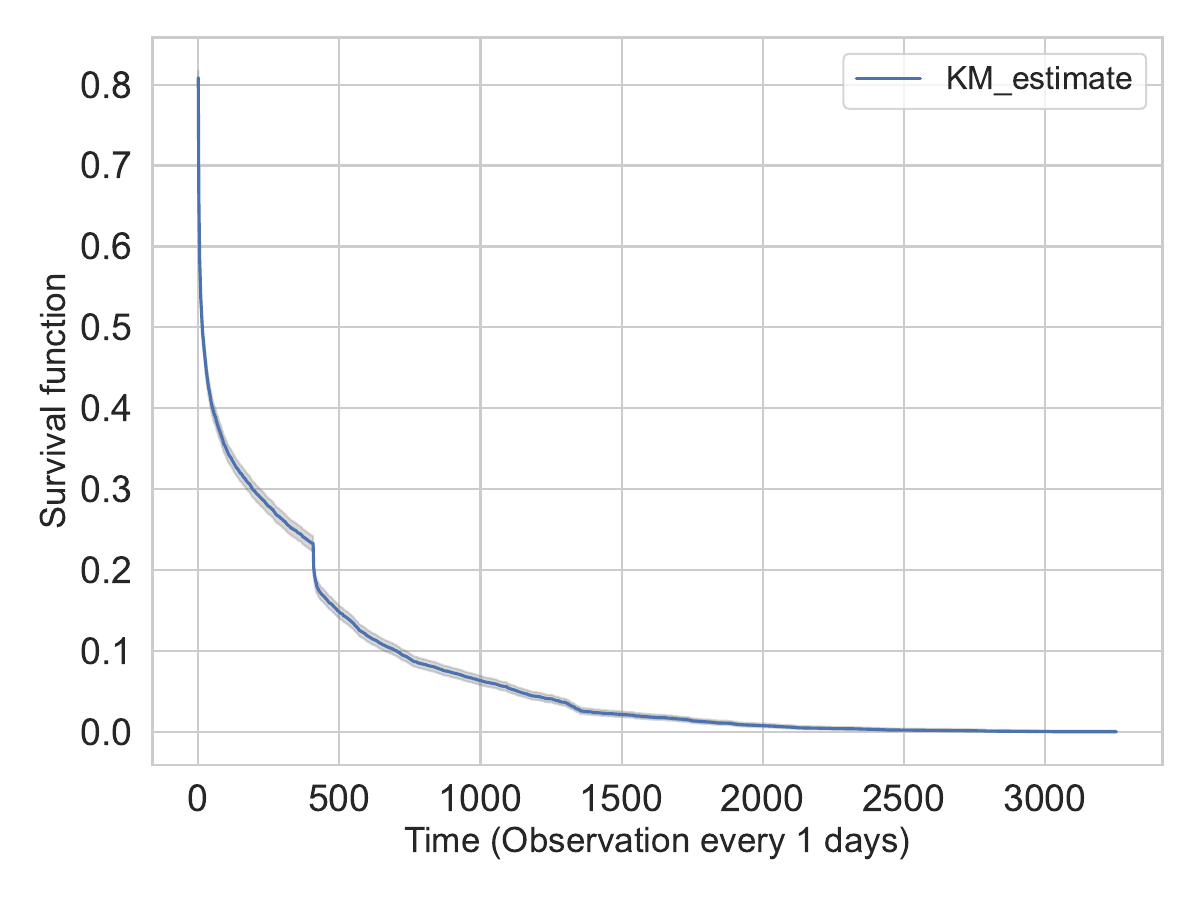}
    \caption{Survival analysis of issue resolution times in Go-Ethereum. The line represents the probability of issue resolution after each day since the issue was opened.}
    \label{fig:survival}
\end{figure}

\paragraph{Characterizing Issue Resolution Through Survival Analysis}
We applied the Kaplan-Meier estimator to analyze resolution times for both open and closed issues, using the last collection date as the endpoint for unresolved issues. Fig.~\ref{fig:survival} shows the survival curve for Go-Ethereum, our most efficient repository, resolving 40\% of issues within days and 75\% within a year. Other repositories show varying patterns: MetaMask and Solidity resolve 50--60\% within a year, while OpenZeppelin and Consensus-Specs match Go-Ethereum's efficiency. Truffle, Ethers.js, and particularly Hardhat show slower resolution times. The anomaly around 400 days in Go-Ethereum suggests systematic closure of backlogged issues.

\begin{figure}
    \centering
    \includegraphics[width=\linewidth]{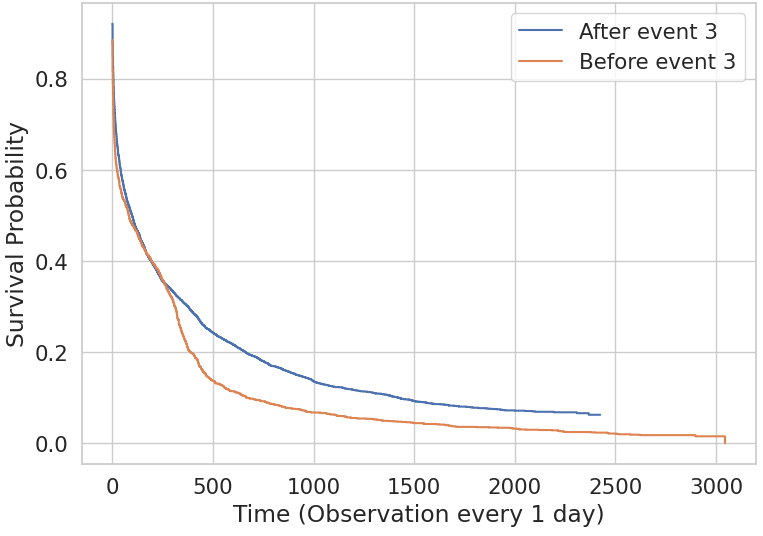}
    \caption{Comparison of issue resolution patterns in MetaMask before and after the Crypto Boom event. The two lines representing issue resolution probabilities before and after the event, can highlight changes.}
    \label{fig:survival-split-4}
\end{figure}

\paragraph{Impact of Events on Survival Analysis}
To assess event influence on issue resolution, we split issues into groups based on their opening time relative to each event and compared their survival curves. Fig.~\ref{fig:survival-split-4} demonstrates this approach using the Crypto Boom (event 3) in MetaMask, where diverging curves indicate event impact on resolution patterns. 
We used log-rank tests to determine statistical significance of these differences, applying the Benjamini-Hochberg procedure across the 10 events for each repository, consistent with our approach in RQ2.
Table~\ref{tab:survival-events} presents the test results across all repositories.

\begin{table*}[h]
\caption{$p$ for log-rank test of survival curves when issues are split into two groups (before and after an event)}
\label{tab:survival-events}
\resizebox{\textwidth}{!}{
\begin{tabular}{lrrrrrrrrrrrr}
\hline
\textit{\textbf{Repository} } & \textbf{1 - Frontier} & \textbf{2 - DAO Hack} &  \textbf{3 - Crypto Boom}  & \textbf{4 - COVID Crash}& \textbf{5 - Beacon} & \textbf{6 - London} & \textbf{7 - Arrow Gl.} & \textbf{8 - Ropsten} & \textbf{9 - Merge} & \textbf{10 - Shanghai}\\
 \hline
\textit{MetaMask} & - & \cellcolor{lightgray}$.249$ & $<.001$ & \cellcolor{lightgray}$.067$ & \cellcolor{lightgray}$.173$ & \cellcolor{lightgray}$.067$ & \cellcolor{lightgray}$.149$ & $<.001$ & $<.001$ & $<.001$\\
\textit{Solidity} & - & \cellcolor{lightgray}$.721$ & $.003$ & \cellcolor{lightgray}$.534$ & \cellcolor{lightgray}$.534$ & \cellcolor{lightgray}$.534$ & \cellcolor{lightgray}$.534$ & \cellcolor{lightgray}$.534$ & \cellcolor{lightgray}$.534$ & \cellcolor{lightgray}$.534$ \\
\textit{Go-ethereum} & $<.001$ & $.003$ & $<.001$ & $<.001$ & $<.001$ & $<.001$ & $<.001$ & $<.001$ & $<.001$ & $<.001$ \\
\textit{Chainlink} & - & - & - & $<.001$ & $<.001$ & $<.001$ & $<.001$ & $<.001$ & $<.001$ & $<.001$ \\
\textit{Truffle} & \cellcolor{lightgray}$.171$ & $<.001$ & $<.001$ & $<.001$ & $<.001$ & $.009$ & $.004$ & $<.001$ & $.002$ & \cellcolor{lightgray}$.218$  \\
\textit{Hardhat} & - & - & - & $<.001$ & $<.001$  & $<.001$ & $<.001$ & $<.001$ & $<.001$ & $<.001$  \\
\textit{Web3-js} & $<.001$ & \cellcolor{lightgray}$.310$  & $<.001$ & $<.001$ & $<.001$ & $<.001$ & $<.001$ & $<.001$ & $<.001$ & $<.001$ \\
\textit{OpenZeppelin} & - & - & \cellcolor{lightgray}$.560$ & \cellcolor{lightgray}$.836$ & \cellcolor{lightgray}$.764$ & \cellcolor{lightgray}$.317$ & $.035$ & $<.001$ & $<.001$ & $<.001$\\
\textit{Consensus-Specs} & - & - & - & $<.001$ & $<.001$ & $<.001$ & $<.001$ & $<.001$ & $<.001$ & $<.001$ \\
\textit{Ethers-js} & - & - & \cellcolor{lightgray}$.344$ & $<.001$ & \cellcolor{lightgray}$.225$ & $.058$ & $.007$ & $<.001$ & $<.001$ & $<.001$ \\

 \hline
\end{tabular}
}
\end{table*}

While many events significantly impacted resolution times across repositories, some repositories like Solidity showed fewer significant changes, suggesting different dynamics in their issue resolution processes. For instance, Go-Ethereum displayed significant differences before and after all the events. MetaMask showed significant changes after the Crypto Boom and the last 3 events.

\begin{tcolorbox}[right=0.1cm,left=0.1cm,top=0.1cm,bottom=0.1cm]
\textbf{Answer to RQ3:} Major Ethereum events impact issue resolution times. Issue creation spikes around key events are often followed by bursts of resolutions as developers prepare for or respond to these changes. While resolution speeds vary across repositories, most issues are resolved quickly, though some repositories appear to manage the number of outstanding issues differently. Overall, issue resolution times change in a statistically significant way before and after major events for most repositories, reflecting increased activity and responsiveness. However, some repositories exhibit resilience to event effects, indicating diverse development dynamics within the ecosystem.
\end{tcolorbox}

\section{Changes in Developer Collaboration Networks During Significant Ethereum Events}\label{RQ4}

We constructed a network to capture interactions within the Ethereum developer community. Each node represents a contributor, and an edge represents a collaboration between two contributors who comment on the same issue. We focus on how collaboration networks change in response to events, so we do not track who created the issue, only who commented on it. Since our goal is to examine collaboration, the issue creator's role is not considered hierarchical, and the networks are undirected.
Across all repositories, we identified 17829 unique comment authors, with 2774 (15\%) contributing to more than one repository. Over half of these overlapping authors (1816) are concentrated in just two repositories. This indicates that each repository largely functions as a distinct network of interactions. Therefore, for this analysis, we treat each repository as an individual, self-contained network.
Let us focus on MetaMask, the largest repository by total activity. We constructed an adjacency matrix where two comment authors are connected if they commented on the same issue, with the connection strength determined by the number of shared issues. This results in an undirected, weighted network of interactions among MetaMask contributors.
The network, spanning from October 22, 2014, to August 28, 2024, consists of 6828 nodes (comment authors), all of whom have collaborated with at least one other contributor. The degree distribution of the network is highly right-skewed, following a long-tailed distribution. Most nodes have a low degree, meaning contributors engage with a small number of others by commenting on a few issues. However, a few nodes exhibit very high degrees, indicating involvement in numerous issues and collaboration with many contributors.
This distribution is typical of a scale-free network, where a few highly connected nodes (hubs) dominate \cite{barabasi2009scale}. These hubs likely represent core contributors who play a key role in MetaMask's development, while the majority of contributors participate sporadically, a common pattern in open-source communities.
Additional metrics further illustrate the network structure: the network density is .0027, indicating sparsity, and the average clustering coefficient is .0003, reflecting low local clustering \cite{newman2012communities}. The network contains 24 connected components, indicating some fragmentation.

\begin{figure}[h]
    \begin{minipage}{0.5\textwidth}
        \centering \includegraphics[width=\textwidth]{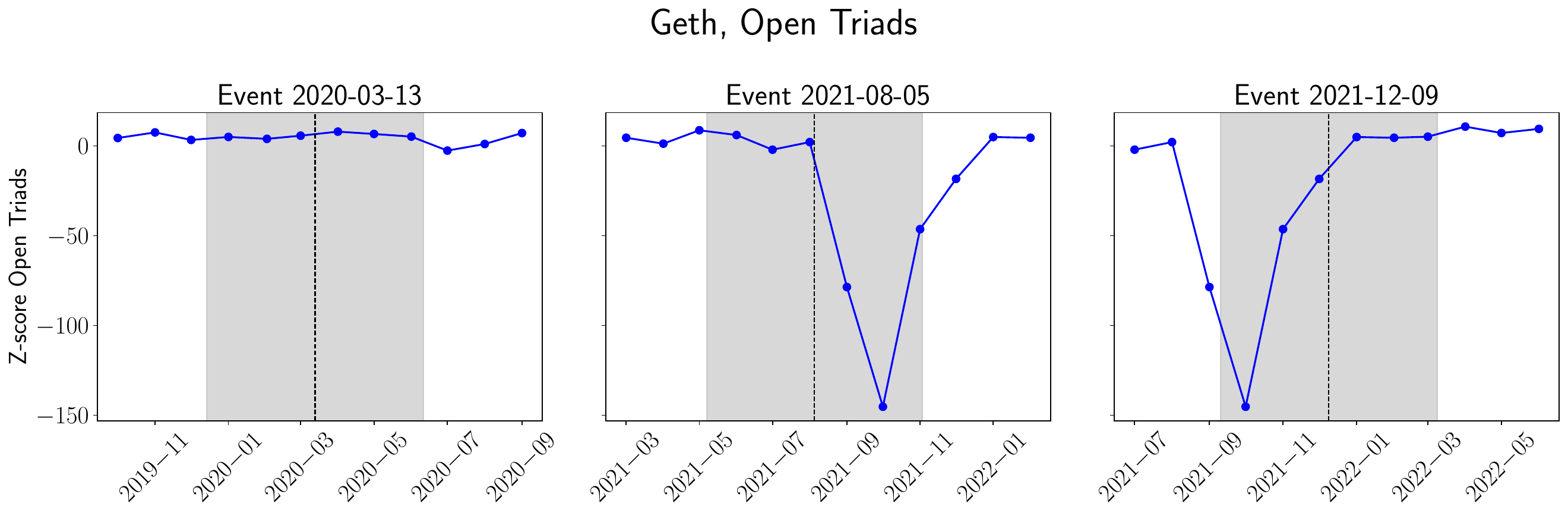} \subcaption{Open tryads}\label{fig: geth open tryads}
    \end{minipage}%
    \hspace{0.05\textwidth}
    \begin{minipage}{0.5\textwidth}
        \centering \includegraphics[width=\textwidth]{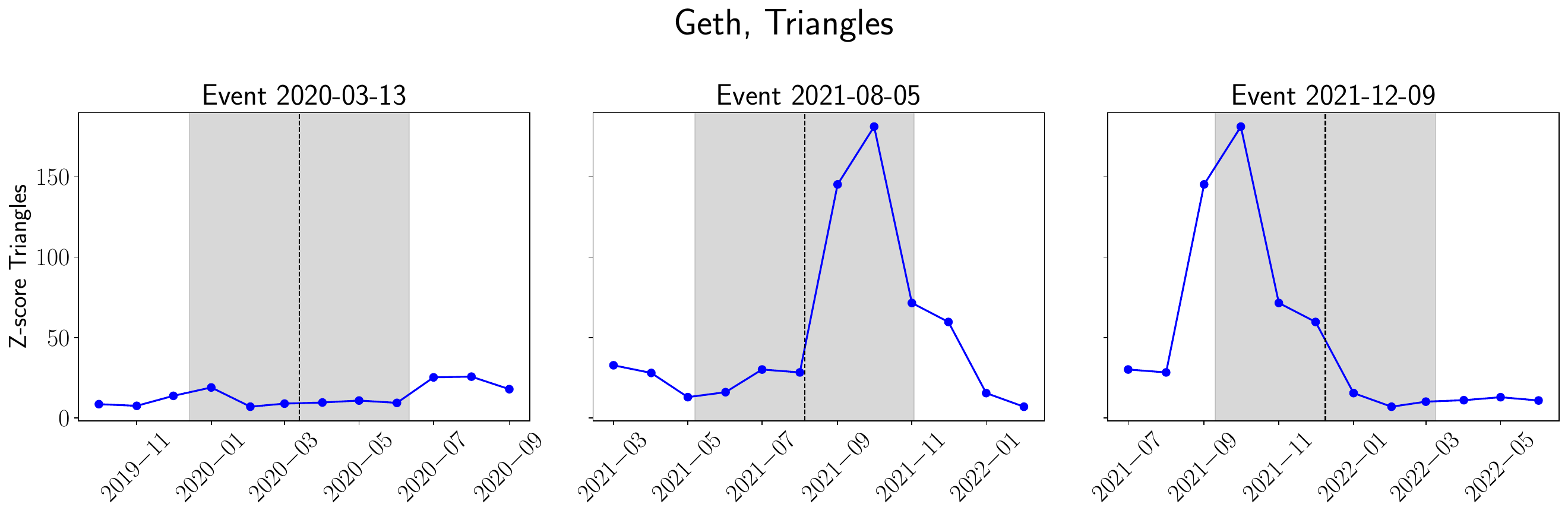} \subcaption{Triangles}\label{fig: geth triangles}
    \end{minipage}
    \caption{Go-ethereum repository. The black vertical line marks the event, while the grey shaded area represents the three months before and after the event. The blue line shows the trend of the Z-score for the triadic motifs over time} \label{fig: geth triadic motifs}
\end{figure}

Following Squartini et al. (2013) \cite{squartini2013early}, we analyze the network evolution by examining topological signatures in monthly intervals from January 26, 2016 to August 2024 (104 months). This monthly resolution aligns with findings from Sec.~\ref{RQ1}, where event impacts manifest within 90-day periods. We represent collaboration as symmetric, weighted networks where nodes are developers and edge weights indicate shared issue comments, yielding monthly $N \times N$ adjacency matrices $A_t$, with $(A_t)_{ij}$ representing interaction strength between developers $i$ and $j$ at month $t$ (where $t = 1, \cdots, 104$).
Beyond network size and density, we analyze higher-order topological properties through motifs, which reveal complex interaction patterns and emerging trends. We examine the relative frequency of undirected network motifs: reciprocated dyads (\(D^{\Leftrightarrow}\)), open triads (V-shape, two nodes connected to a common node), and closed triads (triangles). Using the Configuration Model (CM) as a null model, we calculate Z-scores to quantify deviations from random expectations, comparing each measured quantity $X$ to its expected value $\langle X \rangle$. 
Unlike the Erdős–Rényi model, CM preserves degree distribution while randomizing link weights.
Following findings from Sec.~\ref{RQ2}, we analyze three key events: the COVID-19 crash (March 13, 2020), London Hard Fork (August 5, 2021), and Arrow Glacier Update (December 9, 2021). We examine Z-scores of triadic motifs, which better capture complex collaboration structures than dyadic ones, during six-month windows around each event, with grey highlighting marking the three-month impact period. 
As shown in Fig.~\ref{fig: geth triadic motifs}, Go-ethereum exhibits the strongest patterns, with technical events triggering significant drops in open triads and spikes in triangles. 
While MetaMask and Solidity show similar but muted effects, the COVID-19 crash had minimal impact across repositories compared to technical events' substantial influence on core infrastructure collaboration patterns. The market crash in March 2020 had the least impact on the interaction structure. In contrast, the London Hard Fork and Arrow Glacier Update had more significant effects, particularly on closed triads, suggesting tighter collaboration during technical events. 
Go-Ethereum consistently shows higher Z-scores across both open and closed triads, indicating its role as a central hub during both financial and technical shifts, likely due to its importance in defining smart contracts. MetaMask and Geth were more reactive during market events, reflecting greater sensitivity to market conditions, while Solidity remained stable, showing its consistent role within the ecosystem.

\begin{tcolorbox}[right=0.1cm,left=0.1cm,top=0.1cm,bottom=0.1cm]
\textbf{Answer to RQ4:} The March 2020 market crash had the least impact on interaction structures, while the London Hard Fork and Arrow Glacier Update had a more significant effect, suggesting increased collaboration during technical events. Go-ethereum showed its central role during financial and technical changes, likely due to its importance in smart contract development. MetaMask and Go-ethereum were more reactive to market events, while Solidity remained stable, showing resilience in the ecosystem.
\end{tcolorbox}

\section{Threats to Validity}
\label{threats}

\textbf{Construct Validity:} Primary threats involve our operationalization of key concepts. Our classification of major Ethereum events involves subjectivity, though mitigated by selecting widely acknowledged events with cross-domain impact. Using commit activity and issue resolution time as development metrics may not capture all contributions or team dynamics, while network analysis through shared comments might miss other collaboration forms. We address these limitations through multiple metrics and cross-repository validation. Repository selection bias is minimized by using objective criteria (activity levels, ecosystem roles) to ensure complete coverage of the Ethereum ecosystem.
To ensure broad impact, we define major events as those spanning at least two categories (infrastructure, market, or development trajectory). This criterion avoids overestimating the influence of isolated events. Our classification is based on historical records beyond our dataset, including Ethereum Foundation communications, network upgrades, market shifts, and security reports from 2014–2024. While a constructed definition, it provides consistency in assessing event significance. Future work could refine this approach by exploring alternative classification methods.

\textbf{Internal Validity:} Several factors could affect the relationship between events and observed developer activity. First, concurrent events or changes not included in our analysis might influence our results. We addressed this through our 90-day window analysis and by validating findings against random time periods.
Repository sizes vary significantly (from 678 to 24617 commits). However, our key findings and primary statistical inferences are drawn from the major repositories ($>15K$ commits): Go-ethereum, MetaMask, and Solidity. While we analyze medium ($5-15K$) and smaller ($<5K$) repositories for completeness, their results primarily serve to complement our main conclusions. This approach ensures our statistical inferences remain robust despite the size variations across the dataset.\\
\textbf{External Validity:} Our study focuses specifically on the Ethereum ecosystem during 2014–2024, examining how different types of events impact development patterns across its diverse repositories. While our findings provide insights into Ethereum's development dynamics, we acknowledge that these patterns are shaped by Ethereum's unique characteristics as a leading smart contract platform. Factors such as its decentralized governance, token incentives, and community-driven protocol upgrades differentiate Ethereum from traditional OSS projects.
Our repository selection spans multiple layers of the Ethereum ecosystem—from core infrastructure (Go-ethereum) to development tools (Hardhat, Truffle) and user-facing applications (MetaMask)—providing a broad view of development patterns within this blockchain platform. This diversity strengthens our findings regarding how different components of the Ethereum ecosystem respond to various types of events. However, we recognize that event-response patterns may differ in non-Ethereum projects, where governance models, incentive structures, and development workflows vary. 
Future research could explore whether similar trends hold across other blockchain and non-blockchain OSS ecosystems, particularly in projects with distinct governance mechanisms or without direct market exposure.\\ 
\textbf{Conclusion Validity:} We ensured statistical reliability by selecting appropriate tests based on data distributions, calculating effect sizes, and conducting multiple complementary analyses. Our survival and network analyses are based on assumptions of censoring independence and comment co-occurrence as a proxy for collaboration. To address multiple comparisons, we applied the Benjamini-Hochberg procedure to control the false discovery rate (FDR).  
Data completeness may be influenced by GitHub API limitations, particularly for older events. However, to support validation and reproducibility, we provide a complete replication package containing all data, code, and analysis scripts \cite{Vaccargiu2025}.

\section{Conclusion}
\label{conclusion}

Our analysis of the Ethereum ecosystem offers insights into open-source software development dynamics, particularly in contexts where community-driven development intersects with market pressures. Traditional OSS projects rarely experience such direct market influences, making it challenging to study how economic forces shape development patterns. The Ethereum ecosystem, however, provides a natural experiment where market events, technical upgrades, and community decisions create conditions similar to those faced by commercial software development teams.
The observed distinction between responses to planned technical changes and unexpected market events parallels challenges faced in commercial software development. Through our temporal analysis, we found that event impacts persist for approximately three months before activity patterns normalize, with key events like the Frontier Release and DAO hack showing the strongest effects. Our findings suggest that OSS communities, typically insulated from market pressures, can maintain development momentum through both planned and unplanned changes when proper coordination mechanisms exist.

Network analysis reveals how different types of events influence collaboration patterns, with technical transitions promoting stronger core team interactions and market events driving broader community engagement. This pattern, particularly evident in Go-ethereum's central role during both technical and market events, offers lessons about team structure during different types of organizational changes. 
Core repositories' rapid issue resolution alongside varying patterns in development tools shows OSS communities can balance market pressures with technical excellence, challenging assumptions about market versus community-driven development.

\balance
\bibliographystyle{IEEEtran}
\bibliography{biblio}

\end{document}